\documentclass[11pt,a4,twoside,longbibliography]{article}
\usepackage[dvips,a4paper,left=2.0cm,right=2.0cm,top=2.0cm,bottom=2.0cm,headsep=1em]{geometry}
\usepackage{fancyhdr}
\usepackage{titlesec}
\usepackage[latin1]{inputenc}
\usepackage{amsmath,amsfonts,amssymb,array,amsthm,mathrsfs}
\usepackage{rotating}
\usepackage[font={small}]{caption}
\usepackage{lmodern,slantsc}
\usepackage{multirow}
\usepackage{chngcntr}
\usepackage{placeins}
\usepackage{caption}
\usepackage{enumitem}
\usepackage{cite}
\usepackage[sort&compress,numbers]{natbib}
\def\bibsep{6pt}
\def\bibfont{\small}
\usepackage[breaklinks, bookmarks=true,
colorlinks=true, linkcolor=blue, citecolor=blue, urlcolor=blue, filecolor=blue,
pdfborder={0 0 0}, pdfpagelabels]{hyperref}
\usepackage{tikz}

\let\oldabstract\abstract
\let\oldendabstract\endabstract
\makeatletter
\renewenvironment{abstract}
{%
{\list{}{\addtolength{\leftmargin}{-0.2em} 
\listparindent 1.5em%
\itemindent\listparindent%
\rightmargin\leftmargin%
\parsep\z@\@plus\p@}%
\item\relax}%
{\endlist}%
\oldabstract}
{\oldendabstract}
\makeatother

\def\vr{\mathbf r}

\def\v0{\boldsymbol{0}}

\newlength{\FigureHeight}
\newlength{\FigureHeightHalf}

\numberwithin{equation}{section}
\titleformat{\section}
{\large\bfseries}{\thetitle.}{0.5em}{}
\titleformat{\subsection}
{\normalfont\bfseries}{\thetitle.}{0.5em}{}
\titleformat{\subsubsection}
{\normalfont\itshape}{\normalfont\thetitle.}{0.5em}{}
\begin{document}

\title{\vspace{-1.5em} No new scaling laws of passive scalar with a constant mean gradient in decaying isotropic turbulence}

\author{Michael Frewer\hspace{0.5mm}\thanks{Email address for correspondence:
frewer.science@gmail.com}\\[0.5em]
\small Heidelberg, Germany\\
}
\date{{\small\today}}
\clearpage \maketitle \thispagestyle{empty}

\vspace{-2.0em}\begin{abstract}

\noindent
In the study by Sadeghi \& Oberlack [\href{https://doi.org/10.1017/jfm.2020.413}{JFM~899,~A10~(2020)}] it is claimed that new scaling laws are derived for the case of passive scalar
dynamics under the influence of a constant mean gradient in decaying homogeneous isotropic turbulence. However, these scaling laws are not new and have already been derived and discussed in
\href{http://arks.princeton.edu/ark:/88435/dsp015t34sm99d}{Bahri (2016)}. No novel analytical achievements are made by Sadeghi \& Oberlack, as the title of their study misleadingly wants to suggest. In fact, the already established self-similar scaling laws obtained by Bahri through simple dimensional analysis are already more general in the application than the ones obtained by Sadeghi \& Oberlack through an overly complicated and therefore unnecessarily performed Lie-group symmetry analysis. The~claim that it has the virtue of not being an {\it ad-hoc} method is not true. Because, instead of using an {\it a~priori} set of scales as the classical~method, the Lie-group method has to make use of an {\it a~priori} set of symmetries, namely to select the correct relevant symmetries from an infinite and thus unclosed set. For~example,
a nonphysical scaling symmetry is selected which in the course of the analysis has to be discarded since it is not compatible with the data simulated. Hence, the Lie-group symmetry method in turbulence is just another common trial-and-error method and not a first-principle method that can bypass the closure problem.

\vspace{0.5em}\noindent{\footnotesize{\bf Keywords:} {\it Statistical Physics, Turbulence, Symmetry Analysis, Lie-Groups, Scaling, Closure Problem}}
\end{abstract}

\vspace{0.0em}
\section{Proof that the scaling laws in \texorpdfstring{\cite{Sadeghi20.2}}{Sadeghi20} are not new\label{Sec1}}

To have an overview of what will be proven in this section, here a brief summary:
\begin{itemize}

\item[1.] The first key result (4.24) derived in \cite[Sadeghi \& Oberlack (2020)]{Sadeghi20.2} corresponds exactly to the already published result (4.20) in \cite[Bahri (2016)]{Bahri16},
    or alternatively to the result (2.14)\linebreak[4] in \cite[Bahri {\it et al.} (2015)]{Bahri15}.

\item[2.]  The second key result (4.25) in \cite[Sadeghi \& Oberlack (2020)]{Sadeghi20.2} corresponds exactly to the result~(5.10) in \cite[Bahri (2016)]{Bahri16}. Note that this result was not published
    in \cite[Bahri {\it et al.} (2015)]{Bahri15}\linebreak[4] and thus can only be found in \cite[Bahri (2016)]{Bahri16}, but which is not mentioned or cited in \cite[Sadeghi \& Oberlack (2020)]{Sadeghi20.2}.

\item[3.] The third result (4.26) in \cite[Sadeghi \& Oberlack (2020)]{Sadeghi20.2} also constitutes no new result as claimed, since it is just a direct consequence of Bahri's scaling laws when applied correctly to the underlying equations. The same is true of their results (5.7)-(5.10), which again are just a direct consequence of Bahri's results. In particular, the last result (5.10) was already established back in the 1990s by George \& Speziale {\it et al.}

\item[4.] While \cite[Sadeghi \& Oberlack (2020)]{Sadeghi20.2} considers in their scaling only the case of zero thermal diffusivity~$(\alpha=0)$, Bahri {\it et al.} considered the case of non-zero thermal
    diffusivity~$(\alpha\neq 0)$. Hence, one could assert that the scaling laws (4.24)-(4.26) derived in \cite[Sadeghi \& Oberlack (2020)]{Sadeghi20.2}\linebreak[4] are indeed different and thus can be claimed as novel. But, as will be proven below, such an assertion is invalid. The self-similarity results of Bahri {\it et al.} include and clearly go beyond those of Sadeghi \& Oberlack.

\end{itemize}

\subsection{Proof that (4.24) and (4.25) in \texorpdfstring{\cite{Sadeghi20.2}}{Sadeghi20} are not new\label{Sec1.1}}

The two key results, the temporal scaling law (4.24) and (4.25) in \cite{Sadeghi20.2}
\begin{equation}
\overline{\theta^2}\propto(t+t_0)^{2n+m}, \qquad \overline{u_2\theta}\propto (t+t_0)^{2n+m-1},
\label{200723:1038}
\end{equation}
where $t_0$, $n$ and $m$ are arbitrary constants, correspond exactly to the already published scaling laws~(2.14) in \cite{Bahri15}, and to (4.20) and (5.10) in \cite{Bahri16}, respectively
\begin{equation}
\overline{\theta^2}=C_2[\ell_\theta]^{2C_1\alpha/A_\theta-1},\qquad \overline{v\theta}=C_2\frac{\alpha}{\beta}[\ell_\theta]^{2C_1\alpha/A_\theta-3},
\label{200723:1025}
\end{equation}
where the scalar (temperature) length scale $\ell_\theta$ is given by (2.10) in \cite{Bahri15}, or alternatively by (4.16)\linebreak[4] in~\cite{Bahri16}, as
\begin{equation}
\ell_\theta^2=\ell_{0\theta}^2+A_\theta(t-t_0),
\label{200723:1017}
\end{equation}
where $\ell_{0\theta}$, $A_\theta$ and $t_0$ are arbitrary constants. To note is the typo in \cite{Bahri16}, where the exponent for the scaling law~(5.10) misses an additional $-2$, which can be easily seen when deriving~(5.10) from the correctly printed relations (5.9) and (5.6).

\vspace{1em}\noindent
{\it Proof:} By re-defining the constant virtual time origin $t_0$ in \eqref{200723:1017} as
\begin{equation}
A_\theta t_0^\prime\:\underset{\text{def}}{:=}\:\ell^2_{0\theta}-A_\theta t_0,
\end{equation}
we can equivalently rewrite \eqref{200723:1017} as
\begin{equation}
\ell_\theta^2=A_\theta(t+t_0^\prime),
\end{equation}
and therefore both scaling laws in \eqref{200723:1025} as
\begin{align}
\overline{\theta^2}&=C_2[\ell^2_\theta]^{C_1\alpha/A_\theta-1/2},& \overline{v\theta}&=C_2\frac{\alpha}{\beta}[\ell^2_\theta]^{C_1\alpha/A_\theta-1/2-1}\nonumber\\[0.5em]
&=C_2[A_\theta(t+t_0^\prime)]^{C_1\alpha/A_\theta-1/2}, & &=C_2\frac{\alpha}{\beta}[A_\theta(t+t_0^\prime)]^{C_1\alpha/A_\theta-1/2-1}\nonumber\\[0.5em]
&\propto (t+t_0^\prime)^{C_1\alpha/A_\theta-1/2}, & &\propto (t+t_0^\prime)^{C_1\alpha/A_\theta-1/2-1}.
\label{200723:1027}
\end{align}
If we now re-define the constant exponent in terms of the parameters used in \cite{Sadeghi20.2} as
\begin{equation}
C_1\alpha/A_\theta-1/2\:\underset{\text{def}}{=:}\: 2n+m,
\end{equation}
then we get
\begin{equation}
\overline{\theta^2}\propto(t+t^\prime_0)^{2n+m}, \qquad \overline{v\theta}\propto (t+t^\prime_0)^{2n+m-1},
\label{200723:1550}
\end{equation}
which are exactly the two ``new" scaling laws \eqref{200723:1038} derived in \cite{Sadeghi20.2}, where $v=u_2$ is the cross stream velocity. \qed

\vspace{1em}\noindent
Two things should be noted here:
First, Bahri {\it et al.} derive their scaling laws in spectral space, while Sadeghi \& Oberlack in physical space. But this difference is not the issue here, especially since the scaling laws are purely temporal and thus for~1-point quantities exactly the same in spectral as in physical space. The issue here is that Sadeghi \& Oberlack claim these laws as ``new", which is not~true.

Second, the parameter $m$ in \eqref{200723:1038} in \cite{Sadeghi20.2} results from a nonphysical scaling symmetry which the authors used. Their own analysis, however, later revealed this nonphysical feature when they tried to ensure compatibility with the simulated data, with the result that this symmetry, and thus its associated parameter $m$, must be set to zero (see the last sentence on p.$\,$18 in \cite{Sadeghi20.2}). Therefore, exactly as in \eqref{200723:1027}, the scaling \eqref{200723:1550} due to $m=0$ only exhibits one essential scaling exponent, and not two.

\pagebreak[4]
\subsection{Proof that the self-similar results in \texorpdfstring{\cite{Bahri16,Bahri15}}{Bahri16, Bahri15} are more general than in \texorpdfstring{\cite{Sadeghi20.2}}{Sadeghi20}\label{Sec1.2}}

One could make the assertion that the proof in the previous section is inappropriate for the case studied in \cite{Sadeghi20.2}, for the reason that therein the case of zero thermal diffusivity ($\alpha=0$) was considered,
while Bahri~{\it et~al.} considered the case of non-zero thermal diffusivity ($\alpha\neq0$). Hence, one could assert that the two scaling laws (4.24) and (4.25) in \cite{Sadeghi20.2} are indeed different and thus can be claimed as novel. But, as will be proven now, such an assertion is not valid.

First of all, Bahri {\it et al.} did {\it not} rule out the case $\alpha=0$. Their study is valid for all $\alpha\in\mathbb{R}$,\linebreak[4] and thus includes the special case $\alpha=0$ which Sadeghi \& Oberlack considered. To demonstrate this, let's look more closely, for example,\footnote{All arguments and proofs brought forward here for the scaling law of the temperature variance $\overline{\theta^2}$, equally applies to Bahri's scaling law of the scalar flux $\overline{v\theta}$ in \cite{Bahri16}.} at the scaling law for the temperature variance (2.14) in \cite{Bahri15}
\begin{equation}
\overline{\theta^2}=C_2[\ell_\theta]^{2C_1\alpha/A_\theta-1},
\label{200801:1700}
\end{equation}
which, as was proven before in Sec.$\,$\ref{Sec1.1}, can be rewritten as
\begin{equation}
\overline{\theta^2}\propto(t+t^\prime_0)^{2n+m},
\label{200801:1728}
\end{equation}
where
\begin{equation}
\ell_\theta^2=\ell_{0\theta}^2+A_\theta(t-t_0),\qquad A_\theta t_0^\prime\:\underset{\text{def}}{:=}\:\ell^2_{0\theta}-A_\theta t_0,\qquad
C_1\alpha/A_\theta-1/2\:\underset{\text{def}}{=:}\: 2n+m,
\label{200801:1705}
\end{equation}
with $C_1$, $C_2$, $A_\theta$, $\ell_{0\theta}^2$, and $t_0$ being arbitrary constants.

Now, if we naively put $\alpha=0$ in \eqref{200801:1700}, it will result in a parameter-independent scaling exponent, which, according to \eqref{200801:1705}, will imply the restriction
$2n+m=-1/2$ on the Sadeghi-Oberlack-parameters~$(n,m)$ --- a restriction, however, which is not present in \cite{Sadeghi20.2}, regardless of whether $m$ is zero or not. Naively one can thus conclude that the scaling for $\alpha=0$ in \cite{Sadeghi20.2} is more general than in~\cite{Bahri15}. But this is a fallacy!

There are two ways how to prove that \cite{Sadeghi20.2} is {\it not} more general for $\alpha=0$ than \cite{Bahri15}, where both proofs make use of the fact that the scaling can be made independent of~$\alpha$. Too see that, we have to go back to the derivation of~\eqref{200801:1700} in \cite{Bahri15}. The central terms therein are (2.2)-(2.4) and (2.12). Proof~A~(see~below) directly puts $\alpha=0$ in (2.2) in \cite{Bahri15}, while Proof B separates out $\alpha$ in (2.2) in \cite{Bahri15} by not specifying a particular value for it, thus showing that the scaling laws of Bahri {\it et al.} are valid for {\it all}~$\alpha\in\mathbb{R}$, including the specific case $\alpha=0$.

Hence, the self-similarity results of Bahri {\it et al.} go beyond those of Sadeghi \& Oberlack.\linebreak[4]
In other words, not the results of Sadeghi \& Oberlack, but ultimately the results of Bahri {\it et~al.}\linebreak[4] are the more general ones.

\vspace{1em}
\noindent
{\it Proof A:} Starting off with (2.2) in \cite{Bahri15} and directly putting $\alpha$ to zero, we get:
\begin{equation}
[\dot{E}_{s,\theta}]f_1+\left[E_{s,\theta}\frac{\dot{\ell}_\theta}{\ell_\theta}\right]f_1^\prime\gamma=-[\beta E_{s,v\theta}]f_2-[T_{s,\theta}]g_1.
\end{equation}
Dividing this whole equation by $E_{s,\theta}/\ell^2_\theta$, we then get
\begin{equation}
\left[\dot{E}_{s,\theta}\frac{\ell^2_\theta}{E_{s,\theta}}\right]f_1+\left[\ell_\theta\dot{\ell}_\theta\right]f_1^\prime\gamma=
-\left[\beta \frac{E_{s,v\theta}}{E_{s,\theta}}\ell^2_\theta\right]f_2-\left[\frac{T_{s,\theta}}{E_{s,\theta}}\ell^2_\theta\right]g_1,
\end{equation}
which, if self-similarity should hold, leads to the following conditions
\begin{align}
\left[\dot{E}_{s,\theta}\frac{\ell^2_\theta}{E_{s,\theta}}\right]=\epsilon_1^0,\label{200801:1822}\\[0.5em]
\left[\ell_\theta\dot{\ell}_\theta\right]=\epsilon_2^0,\label{200801:1834}\\[0.5em]
\left[\beta \frac{E_{s,v\theta}}{E_{s,\theta}}\ell^2_\theta\right]=\epsilon_3^0,\\[0.5em]
\left[\frac{T_{s,\theta}}{E_{s,\theta}}\ell^2_\theta\right]=\epsilon_4^0,\label{200801:1823}
\end{align}
where $\epsilon_1^0$, $\epsilon_2^0$, $\epsilon_3^0$, and $\epsilon_4^0$ are arbitrary constants without any restrictions, with the 0-index indicating that $\alpha=0$. The above conditions \eqref{200801:1822}-\eqref{200801:1823} are the same as (2.3)-(2.6) in~\cite{Bahri15}, yet, only independent of $\alpha$. Hence, when renaming the first two $\epsilon$-constants\hspace{0.5mm}\footnote{The renaming of $\epsilon_2$ in \cite{Bahri15} above (2.10) has a typo, which easily can be verified when integrating (2.4) to~(2.10) in~\cite{Bahri15}: Instead of `$\epsilon_2=\alpha A_\theta$' it should read `$\epsilon_2=\alpha A_\theta/2$'.}
\begin{equation}
\epsilon_1^0=C_1^0, \qquad \epsilon_2^0=A^0_\theta/2,
\label{200801:1947}
\end{equation}
as in \cite{Bahri15}, but now indicating that $\alpha=0$, then solving for \eqref{200801:1822} and \eqref{200801:1834}, and finally using the relation (2.12) in \cite{Bahri15}, we end up with the same scaling law (2.14) as before,
\begin{equation}
\overline{\theta^2}=C_2[\ell_\theta]^{2C^0_1/A^0_\theta-1},
\label{200801:1837}
\end{equation}
but, now, only independent of $\alpha$. Result \eqref{200801:1837} shows that for $\alpha=0$ the scaling exponent is indeed parameter-dependent, thus implying the identity  $C^0_1/A^0_\theta-1/2=2n+m$ without any restrictions on the Sadeghi-Oberlack-parameters~$(n,m)$.\qed

\vspace{1em}
\noindent
{\it  Proof B:} We start again with (2.2) in \cite{Bahri15}, but now we do not specify any value for $\alpha$, but leave it arbitrary, i.e., $\alpha\in\mathbb{R}$ (including thus also the case $\alpha=0$):
\begin{equation}
[\dot{E}_{s,\theta}]f_1+\left[E_{s,\theta}\frac{\dot{\ell}_\theta}{\ell_\theta}\right]f_1^\prime\gamma=-[\beta E_{s,v\theta}]f_2-[T_{s,\theta}]g_1-\left[\alpha\frac{E_{s,\theta}}{\ell^2_\theta}\right]2\gamma^2f_1,
\quad \alpha\in\mathbb{R}.
\end{equation}
Dividing this whole equation again by $E_{s,\theta}/\ell^2_\theta$, we then get
\begin{equation}
\left[\dot{E}_{s,\theta}\frac{\ell^2_\theta}{E_{s,\theta}}\right]f_1+\left[\ell_\theta\dot{\ell}_\theta\right]f_1^\prime\gamma=
-\left[\beta \frac{E_{s,v\theta}}{E_{s,\theta}}\ell^2_\theta\right]f_2-\left[\frac{T_{s,\theta}}{E_{s,\theta}}\ell^2_\theta\right]g_1
-\alpha\cdot 2\gamma^2f_1,\quad \alpha\in\mathbb{R},
\label{200801:1945}
\end{equation}
which, if self-similarity should hold, leads to the following conditions
\begin{align}
\left[\dot{E}_{s,\theta}\frac{\ell^2_\theta}{E_{s,\theta}}\right]=\epsilon^\alpha_1,\label{200801:2001}\\[0.5em]
\left[\ell_\theta\dot{\ell}_\theta\right]=\epsilon^\alpha_2,\label{200801:2002}\\[0.5em]
\left[\beta \frac{E_{s,v\theta}}{E_{s,\theta}}\ell^2_\theta\right]=\epsilon^\alpha_3,\\[0.5em]
\left[\frac{T_{s,\theta}}{E_{s,\theta}}\ell^2_\theta\right]=\epsilon^\alpha_4,\label{200801:2003}
\end{align}
where $\epsilon^\alpha_1$, $\epsilon^\alpha_2$, $\epsilon^\alpha_3$, and $\epsilon^\alpha_4$ are arbitrary constants without any restrictions, with the $\alpha$-index now indicating that $\alpha$ can take any value in the defining equation \eqref{200801:1945}, including\pagebreak[4] the value~$\alpha=0$. If~we now rename the first two $\epsilon$-constants again as before in \eqref{200801:1947}, but now indicating these as being constants for arbitrary~$\alpha$,
\begin{equation}
\epsilon_1^\alpha=C_1^\alpha, \qquad \epsilon_2^\alpha=A^\alpha_\theta/2,
\label{200801:1948}
\end{equation}
and then solving the corresponding equations \eqref{200801:2001}-\eqref{200801:2002}, we end up again with the very same scaling law as given in \eqref{200801:1837}, yet, now for any value of $\alpha$:
\begin{equation}
\overline{\theta^2}=C_2[\ell_\theta]^{2C_1^\alpha/A^\alpha_\theta-1},\quad \alpha\in\mathbb{R}.
\label{200801:1951}
\end{equation}
Result \eqref{200801:1951} shows that the scaling structure obtained in \eqref{200801:1837} is valid not only for $\alpha=0$, but for all $\alpha\in\mathbb{R}$. Hence, the scaling result obtained by Bahri {\it et al.} is more general than the result obtained by Sadeghi \& Oberlack, who considered only the case $\alpha=0$.\qed

\subsection{Proof that (4.26) in \texorpdfstring{\cite{Sadeghi20.2}}{Sadeghi20} is not new\label{Sec1.3}}

Also result (4.26) in \cite{Sadeghi20.2} is not a new result as claimed, but only a consequence of Bahri's results \eqref{200723:1550} when correctly applied to the underlying equations. The relevant equations are the equation for the cross stream velocity (3.5) in \cite{Sadeghi20.2}
\begin{equation}
\frac{\partial u_2}{\partial t} + \cdots =0,
\label{200723:1158}
\end{equation}
and the passive scalar (temperature) equation (3.4)
\begin{equation}
\frac{\partial \theta}{\partial t} +u_2\Gamma+\cdots=0,
\label{200723:1159}
\end{equation}
where the dots indicate the remaining terms of the equations, and where $\Gamma$ is the constant mean scalar gradient. Multiplying \eqref{200723:1158} by $\theta$, \eqref{200723:1159} by $u_2$, and then adding both equations, we get the resulting equation
\begin{equation}
\frac{\partial u_2\theta}{\partial t}+u_2^2\Gamma+\cdots=0,
\end{equation}
which then in statistically averaged form reads
\begin{equation}
\frac{\partial \overline{u_2\theta}}{\partial t}+\Gamma\cdot \overline{u_2^2}+\cdots=0.
\end{equation}
Now, when $\overline{u_2\theta}$ evolves according to Bahri's scaling law \eqref{200723:1550}, then the evolution of the transverse fluctuation variance $\overline{u_2^2}$ is dictated by the above equation as
\begin{equation}
\overline{u_2^2}\:\propto\: \frac{\partial}{\partial t}\overline{u_2\theta}\:\propto\: \frac{d}{dt}(t+t^\prime_0)^{2n+m-1}\propto(t+t^\prime_0)^{2n+m-2},
\label{200723:1739}
\end{equation}
which exactly is the result (4.26) in \cite{Sadeghi20.2}.\qed

\subsection{Proof that (5.7)-(5.10) in \texorpdfstring{\cite{Sadeghi20.2}}{Sadeghi20} are not new\label{Sec1.4}}

Also the 2-point results (5.7)-(5.10) in \cite{Sadeghi20.2} do not constitute any new results, since also those are a direct consequence of Bahri's 1-point results \eqref{200723:1550}. To demonstrate this, we take, for example, the considered 2-point equation~(3.18) in \cite{Sadeghi20.2} (the proof for $R_{\theta\theta}$ and $R_{2\theta}$ is analogous)
\begin{equation}
\frac{\partial R_{22}}{\partial t}+\cdots=0.
\label{200723:1614}
\end{equation}
Now, since we are on a search for temporal scaling laws, we redefine the 2-point function $R_{22}$ by factoring out the corresponding 1-point temporal scaling law \eqref{200723:1739} of Bahri {\it et al.}:
\begin{equation}
R_{22}(t,\vr)\:\underset{\text{def}}{=:}\: \overline{u_2^2}\cdot \tilde{R}_{22}(\tilde{\vr}),
\end{equation}
where we make the ansatz of a self-similar spatial variable $\tilde{\vr}:=\vr/\xi(t)$. Inserting this ansatz into~\eqref{200723:1614}, we get:
\begin{align}
0&=\frac{d \overline{u_2^2}}{dt}\cdot \tilde{R}_{22}+ \overline{u_2^2}\cdot\frac{\partial\tilde{R}_{22}}{\partial t}+\cdots\nonumber\\[0.5em]
&\propto (2n+m-2)(t+t^\prime_0)^{2n+m-3}\cdot\tilde{R}_{22}+(t+t^\prime_0)^{2n+m-2}\frac{\partial \tilde{r}^k}{\partial t}\frac{\partial \tilde{R}_{22}}{\partial \tilde{r}^k}
+\cdots\nonumber\\[0.5em]
&\quad\;= (2n+m-2)(t+t^\prime_0)^{2n+m-3}\cdot\tilde{R}_{22}-(t+t^\prime_0)^{2n+m-2}\frac{r^k}{\xi^2}\frac{d\xi}{dt}\frac{\partial \tilde{R}_{22}}{\partial \tilde{r}^k}+\cdots\nonumber\\[0.5em]
&\quad\;= (2n+m-2)(t+t^\prime_0)^{2n+m-3}\cdot\tilde{R}_{22}-\frac{(t+t^\prime_0)^{2n+m-2}}{\xi}\frac{d\xi}{dt}\cdot\tilde{r}^k\frac{\partial \tilde{R}_{22}}{\partial \tilde{r}^k}+\cdots
\label{200723:1636}
\end{align}
which, in order to achieve a $t$-independent invariant equation, implies the relation
\begin{equation}
\frac{d\xi}{d t}=\gamma\cdot\frac{\xi}{t+t^\prime_0},
\end{equation}
where $\gamma$ is some arbitrary proportionality constant. Hence, we get the solution
\begin{equation}
\xi\propto (t+t^\prime_0)^\gamma,
\end{equation}
which then yields the self-similar variable (5.10) in \cite{Sadeghi20.2}
\begin{equation}
\tilde{\vr}=\frac{\vr}{(t+t^\prime_0)^n},
\end{equation}
if $\gamma$ is particularly specified as $n$, as was done in \cite{Sadeghi20.2} in choosing a particular symmetry configuration. \qed

\vspace{1em}\noindent
Hence, only by using the results of Bahri {\it et al.} \cite{Bahri16,Bahri15} and simple dimensional analysis, we yield the very same results as Sadeghi \& Oberlack in \cite{Sadeghi20.2}, thus showing that their results are not novel as claimed.

\section{Further remarks and points for correction in \texorpdfstring{\cite{Sadeghi20.2}}{Sadeghi20}\label{Sec2}}

{\bf 1.}
The structure of Fig.$\,$1 in \cite{Sadeghi20.2} is a repetition of Fig.$\,$4.1 in \cite{Bahri16} and Fig.$\,$1 in \cite{Bahri15},
only displayed for new DNS data. However, since \cite{Bahri15} in particular is not mentioned or cited at this position
or in the corresponding text on p.$\,$18,\footnote{To note is that although \cite{Bahri15} gets cited in \cite{Sadeghi20.2} for other findings and discussions, it is not cited correctly at the key positions in \cite{Sadeghi20.2}.} the reader gets again the misleading impression that this figure and its power-law fits are unique to \cite{Sadeghi20.2}.
The reader is not made aware of the fact that Bahri {\it et al.} achieved power-law fits of similar quality first, and also that they were the first to recognize
the fact that the scalar variance $\overline{\theta^2}$ is not decaying but growing in time.

\vspace{1em}\noindent
{\bf 2.}
In \cite{Sadeghi20.2} it is stated that {\it ``the Lie symmetry yields the interrelation among the various scaling\linebreak[4] exponents, as it leads to a connection between the temporal scaling of the velocity and scalar\linebreak[4] variance"}~[p.$\,$15]. This result, however, is not a merit of Lie symmetry groups. C.~Bahri got this result first without using Lie-groups, and in the process also obtained more information than Lie-groups could give. Her result (5.10) in \cite{Bahri16} for the temporal scaling of the scalar (heat) flux $\overline{v\theta}$ leads to the conclusion {\it ``that the heat flux is governed only by parameters related to the temperature field. Thus, no information is needed about the velocity field to determine the heat flux in this particular flow configuration. This is of major importance to the turbulence community, and particularly the atmospheric sciences where the scalar flux is of main interest"} [p.$\,$78]. This important information cannot be obtained by the Lie-group method as performed by Sadeghi \& Oberlack in \cite{Sadeghi20.2}, which clearly is a serious drawback compared to the successful method used by Bahri {\it et al.}

\pagebreak[4]\noindent
{\bf 3.}
In \cite{Sadeghi20.2} it is said that in {\it ``Bahri et al. 2015 it has been suggested that the TPC [2-point correlation] may not scale on the integral length scale, but instead on the Kolmogorov length scale, the Taylor microscale or rather the scalar Taylor microscale"} [p.$\,$20]. This statement is misleading, since Bahri {\it et al.} \cite{Bahri15}\linebreak[4] considers 2-point correlations in spectral space and not in physical space as Sadeghi \& Oberlack\linebreak[4] in~\cite{Sadeghi20.2}. In clear contrast to 1-point quantities, the 2-point quantities for the considered flow configuration may show a different sensitivity on the scaling behaviour in spectral space than in physical space.\linebreak[4] Furthermore, Bahri {\it et al.} did~not consider the velocity integral scale in their analysis, as it was done in~\cite{Sadeghi20.2}, but instead considered only the scalar integral scale at that point. Hence, the 2-point correlation analysis in~\cite{Bahri15} cannot be compared with the one done in \cite{Sadeghi20.2}.

\vspace{1em}\noindent
{\bf 4.}
In Fig.$\,$4 in \cite{Sadeghi20.2} it is not clear why for the velocity correlations only $\tilde{R}_{22}$ is displayed. What about the other 2-point velocity correlations $\tilde{R}_{11}$, $\tilde{R}_{33}$, $\tilde{R}_{12}$, etc.? Were they just omitted because they maybe don't show such a good collapse when using the velocity integral length scale?

\vspace{1em}\noindent
{\bf 5.}
The term $2\Gamma R_{2\theta}$ in (3.13) in \cite{Sadeghi20.2} is not correct, since $R_{2\theta}\neq R_{\theta 2}$. It either has to be replaced by~$\Gamma (R_{2\theta}(\vr)+ R_{\theta 2}(\vr))$, or by
$\Gamma (R_{2\theta}(\vr)+ R_{2\theta}(-\vr))$.

\vspace{1em}\noindent
{\bf 6.}
The last remark is on the usefulness of a Lie-group symmetry analysis in turbulence as it is carried out in \cite{Sadeghi20.2}. As was proven in the previous sections, a Lie-group symmetry analysis is only overly complicated and thus unnecessary for the flow configuration considered in \cite{Sadeghi20.2}, since a classical self-similarity analysis already suffices to yield the same results. The classical approach in \cite{Bahri15,Bahri16} even led to more general results than the ones obtained in \cite{Sadeghi20.2}.

But also for other turbulent flow configurations, the Lie-group symmetry method is of no significant analytical relevance as long as the equations are not modelled and remain unclosed~\cite{Frewer22}.
The problem here is that we have turbulence, which unfortunately comes along with unclosed statistical equations. Hence, the set of symmetries is also unclosed, which means that if modelling of the statistical equations is not considered as in \cite{Sadeghi20.2}, then (nearly) any symmetry can be generated, and thus also (nearly) any desirable scaling law. The simple reason for this is that at each order of the infinite hierarchy almost any change due to a variable transformation can always be balanced or compensated by an unclosed term at the next higher order. Ultimately this means that the choice of an invariance is made by the user and not dictated by theory, simply because one has an infinite set of invariant possibilities to choose from when performing a full and correct Lie-group symmetry analysis for unclosed equations, as explicitly shown and discussed e.g. in \cite{Frewer22,Frewer21.4,Frewer18.2,Frewer16.3,Frewer14.2}. A crucial information which is not shared with the reader in \cite{Sadeghi20.2}. Instead, as a deflection of the issue, false claims are made regarding the use of Lie-group symmetry method in turbulence, thus raising false hopes, like the following statement on p.$\,$3 in \cite{Sadeghi20.2}:

\vspace{0.75em}\noindent
\emph{``Despite the fact that the classical self-similarity to construct scaling
parameters for the passive scalars has been known for a long time, there are still
several open challenges. First and foremost, the classical self-similarity hypothesis can
only be carried out when using an a priori set of similarity scales for all of the
statistical moments in the transport equations. Therefore, it is one of the outstanding
goals to obtain the similarity scales based on a more general analytical approach rather
than in an ad hoc manner. In addition, the interaction between the velocity and passive
scalar (temperature) fields becomes important in several cases, for example, when a
constant mean temperature gradient is present. However, the lack of a unified approach
that can predict the direct link between the scaling laws for the passive scalar and
velocity moments is noted in the literature.
Therefore, it is the main aim of this paper to use a more general technique, which
is known as Lie symmetry analysis, to formally derive the scaling laws and similarity
variables for a passive scalar flow advected by a turbulent~flow."}

\vspace{0.75em}\noindent
Or, more compact in their abstract: \emph{``It is
shown that, in contrast to the classical self-similarity approach, the general invariant
solutions, respectively scaling laws, of the two-point functions are constructed using
the symmetry approach, without requiring an a priori set of similarity scales to carry
on the analysis."}

\vspace{0.75em}\noindent
Considering the proofs in Sec.$\,$\ref{Sec1}, which demonstrate that a simple dimensional analysis is already fully\linebreak[4] 
sufficient\hfill to\hfill confront\hfill the\hfill current\hfill scaling\hfill problem,\hfill as\hfill correctly\hfill derived\hfill by\hfill Bahri\hfill \emph{et\hfill al.}\hfill in\hfill \cite{Bahri15,Bahri16},\hfill and

\newgeometry{left=2.0cm,right=2.0cm,top=2.0cm,bottom=1.45cm,headsep=1em}

\noindent
the fact that the set of symmetries in turbulence is unclosed, it is easy to unmask the above claims by Sadeghi \& Oberlack as false. It is not true, that the Lie-group symmetry method in turbulence is free of assumptions. It is an {\it ad-hoc} method~too, not in the same but in a similar way as the classical self-similarity method: Instead of using an {\it a priori} set of scales, the Lie-group method has to make use of an {\it a priori} set of symmetries, namely to select the correct relevant symmetries from an infinite~(unclosed) set. In other words, the particular selection of the chosen symmetries (4.7) in \cite{Sadeghi20.2} is a plain assumption, because when performing a full and correct Lie-group symmetry analysis on the considered (unclosed) equations (3.16)-(3.18), one gets an infinite set of possible symmetries,\footnote{To note is that the implied linearity of the equations (3.16)-(3.18) in \cite{Sadeghi20.2} still amplifies the closure problem when correctly applying a Lie-group invariance analysis to it. It is the unpleasant effect of the linear superposition principle which adds an additional infinite dimension to the already infinite dimensional Lie-algebra of invariant transformations that already results from the unclosedness of the equations themselves.} which is not mentioned in \cite{Sadeghi20.2}.

Another problematic issue not mentioned in \cite{Sadeghi20.2} is the fact that due to the arbitrariness involved when making a particular choice from an infinite (unclosed) set of symmetries, there is the high chance that one will select a nonphysical symmetry which is not reflected by experiment or numerical simulation. This clearly is the case for the chosen statistical scaling symmetry (4.6) in \cite{Sadeghi20.2}, with the group parameter $a_s$, and later $m$ (4.20), which clearly is nonphysical --- see e.g. \cite{Frewer22,Frewer17,Frewer16.4,Frewer16.2,Frewer16.1,Frewer15.2,Frewer15.1,Frewer14.1}.

Although this nonphysical symmetry (4.6) later had to be put to zero in \cite{Sadeghi20.2} once a specific initial condition with a prescribed energy spectrum was chosen for the simulation (see p.$\,$18 in \cite{Sadeghi20.2} where the case $m=0$ is discussed), its nonphysical feature is still downplayed as being only a symmetry where {\it ``We do not have an indication from the DNS data that the statistical symmetry (4.6a-n) connected to~$m$ plays a role here. Hence, in the following we set $m=0$"}~[p.$\,$18].

Despite all these signs and warnings, this nonphysical symmetry (4.6) keeps being propagated in \cite{Sadeghi20.2} as a scaling symmetry that plays \emph{``a key role for the understanding of moment scalings in wall-bounded shear flows"} [p.$\,$13], and that \emph{``this symmetry is a measure of intermittency"}~[p.$\,$8]. In particular the latter statement about intermittency,\footnote{To note is that the group Oberlack~{\it et~al.}~use the term  ``intermittency" for the nonphysical scaling symmetry\linebreak[4] interchangeably either for external (large-scale) or internal (small-scale) intermittency, e.g. as done for jet flow in~\cite{Sadeghi21}, or for the transition to turbulence in \cite{Oberlack16}, or for the center region of fully developed channel flow in \cite{Oberlack22}. For the latter case, see also the discussion in \cite{Frewer22}.} here the authors should ask themselves why this symmetry was forced to zero in \cite{Sadeghi20.2} for a flow configuration that exhibits intermittency, regardless of the initial condition considered. In other words, if this scaling symmetry (4.6) truly is a measure of intermittency, then the parameter $m$ for the currently considered flow configuration should be non-zero,~or?

Independent of this apparent conflict, the connection to intermittency of this scaling symmetry~(4.6) is not true and has been clearly refuted several times by now within different statistical frameworks --- see e.g. \cite{Frewer22,Frewer17,Frewer16.4,Frewer16.2,Frewer16.1,Frewer15.2,Frewer15.1,Frewer14.1}. Also from a pure phenomenological viewpoint, it is abundantly clear that intermittency is a symmetry-breaking phenomenon and not a symmetry-existing or symmetry-preserving one. Even if we would wrongly assume this to be the case, intermittency is definitely not described or featured by any global scaling symmetry, particularly not by the one given by (4.6) in~\cite{Sadeghi20.2}, which actually just mimics the standard scaling of a linear system when applied to an unclosed nonlinear system. Obviously, such a symmetry is only a mathematical artefact of the unclosed system itself and therefore indeed nonphysical, independent of the fact, of course, that this symmetry also violates the classical principle of cause and effect \cite{Frewer17,Frewer16.1,Frewer15.1}.

Finally it needs to be noted that in various other publications by Oberlack {\it et~al.}, regarding Lie-group symmetries and turbulence, a second nonphysical symmetry is constantly included, the nonphysical statistical translation symmetry, which led to serious errors, e.g.~in \cite{Avsarkisov14} and~\cite{Sadeghi18}, which then led to the forced Corrigenda \cite{Avsarkisov21} and \cite{Sadeghi20.1}, respectively.
Although \cite{Avsarkisov21}~is a Corrigendum, it is still seriously in error --- for a detailed discussion on this matter, see Appendix~D in \cite{Frewer22}. Also in \cite{Sadeghi20.1}, although a correction
is given, it does not provide an explanation or a correction to the originally published Fig.$\,$7, a figure in \cite{Sadeghi18} that still cannot be reproduced from the data provided. Instead, a completely new figure based on new results is presented and therefore unrelated to the original one. Hence, in both cases \cite{Avsarkisov21} and \cite{Sadeghi20.1}, the same central question: How and with what tools did the authors manage to do the original Fig.$\,$9(a) in \cite{Avsarkisov14} and Fig.$\,$7 in \cite{Sadeghi18}? Nevertheless, although this nonphysical translation symmetry is not included in \cite{Sadeghi20.2}, it can, however, be found again in several recent studies by Oberlack~{\it et~al.}

\restoregeometry

\nocite{apsrev42Control}
\bibliographystyle{apsrev4-2}
\bibliography{References}

\begin{thebibliography}{22}%
\makeatletter
\providecommand \@ifxundefined [1]{%
 \@ifx{#1\undefined}
}%
\providecommand \@ifnum [1]{%
 \ifnum #1\expandafter \@firstoftwo
 \else \expandafter \@secondoftwo
 \fi
}%
\providecommand \@ifx [1]{%
 \ifx #1\expandafter \@firstoftwo
 \else \expandafter \@secondoftwo
 \fi
}%
\providecommand \natexlab [1]{#1}%
\providecommand \enquote  [1]{``#1''}%
\providecommand \bibnamefont  [1]{#1}%
\providecommand \bibfnamefont [1]{#1}%
\providecommand \citenamefont [1]{#1}%
\providecommand \href@noop [0]{\@secondoftwo}%
\providecommand \href [0]{\begingroup \@sanitize@url \@href}%
\providecommand \@href[1]{\@@startlink{#1}\@@href}%
\providecommand \@@href[1]{\endgroup#1\@@endlink}%
\providecommand \@sanitize@url [0]{\catcode `\\12\catcode `\$12\catcode
  `\&12\catcode `\#12\catcode `\^12\catcode `\_12\catcode `\%12\relax}%
\providecommand \@@startlink[1]{}%
\providecommand \@@endlink[0]{}%
\providecommand \url  [0]{\begingroup\@sanitize@url \@url }%
\providecommand \@url [1]{\endgroup\@href {#1}{\urlprefix }}%
\providecommand \urlprefix  [0]{URL }%
\providecommand \Eprint [0]{\href }%
\providecommand \doibase [0]{https://doi.org/}%
\providecommand \selectlanguage [0]{\@gobble}%
\providecommand \bibinfo  [0]{\@secondoftwo}%
\providecommand \bibfield  [0]{\@secondoftwo}%
\providecommand \translation [1]{[#1]}%
\providecommand \BibitemOpen [0]{}%
\providecommand \bibitemStop [0]{}%
\providecommand \bibitemNoStop [0]{.\EOS\space}%
\providecommand \EOS [0]{\spacefactor3000\relax}%
\providecommand \BibitemShut  [1]{\csname bibitem#1\endcsname}%
\let\auto@bib@innerbib\@empty
\bibitem [{\citenamefont {Sadeghi}\ and\ \citenamefont
  {Oberlack}(2020)}]{Sadeghi20.2}%
  \BibitemOpen
  \bibfield  {author} {\bibinfo {author} {\bibfnamefont {H.}~\bibnamefont
  {Sadeghi}}\ and\ \bibinfo {author} {\bibfnamefont {M.}~\bibnamefont
  {Oberlack}},\ }\bibfield  {title} {\bibinfo {title} {New scaling laws of
  passive scalar with a constant mean gradient in decaying isotropic
  turbulence},\ }\href@noop {} {\bibfield  {journal} {\bibinfo  {journal}
  {\href{https://doi.org/10.1017/jfm.2020.413}{J.~Fluid~Mech.}}\ }\textbf
  {\bibinfo {volume} {899}},\ \bibinfo {pages} {A10} (\bibinfo {year}
  {2020})}\BibitemShut {NoStop}%
\bibitem [{\citenamefont {Bahri}(2016)}]{Bahri16}%
  \BibitemOpen
  \bibfield  {author} {\bibinfo {author} {\bibfnamefont {C.}~\bibnamefont
  {Bahri}},\ }\emph {\bibinfo {title} {Fundamentals and scaling of passive
  scalar fields in isotropic turbulence}},\ \href@noop {} {Ph.D. thesis},\
  \bibinfo  {school}
  {\href{http://arks.princeton.edu/ark:/88435/dsp015t34sm99d}{Princeton
  University}} (\bibinfo {year} {2016})\BibitemShut {NoStop}%
\bibitem [{\citenamefont {Bahri}\ \emph {et~al.}(2015)\citenamefont {Bahri},
  \citenamefont {Arwatz}, \citenamefont {George}, \citenamefont {Mueller},\
  and\ \citenamefont {Hultmark}}]{Bahri15}%
  \BibitemOpen
  \bibfield  {author} {\bibinfo {author} {\bibfnamefont {C.}~\bibnamefont
  {Bahri}}, \bibinfo {author} {\bibfnamefont {G.}~\bibnamefont {Arwatz}},
  \bibinfo {author} {\bibfnamefont {W.~K.}\ \bibnamefont {George}}, \bibinfo
  {author} {\bibfnamefont {M.~E.}\ \bibnamefont {Mueller}},\ and\ \bibinfo
  {author} {\bibfnamefont {M.}~\bibnamefont {Hultmark}},\ }\bibfield  {title}
  {\bibinfo {title} {Self-similarity of passive scalar flow in grid turbulence
  with a mean cross-stream gradient},\ }\href@noop {} {\bibfield  {journal}
  {\bibinfo  {journal}
  {\href{https://doi.org/10.1017/jfm.2015.439}{J.~Fluid~Mech.}}\ }\textbf
  {\bibinfo {volume} {780}},\ \bibinfo {pages} {215} (\bibinfo {year}
  {2015})}\BibitemShut {NoStop}%
\bibitem [{\citenamefont {Frewer}\ and\ \citenamefont
  {Khujadze}(2022)}]{Frewer22}%
  \BibitemOpen
  \bibfield  {author} {\bibinfo {author} {\bibfnamefont {M.}~\bibnamefont
  {Frewer}}\ and\ \bibinfo {author} {\bibfnamefont {G.}~\bibnamefont
  {Khujadze}},\ }\bibfield  {title} {\bibinfo {title} {A closer look at
  predicting turbulence statistics of arbitrary moments when based on a
  non-modelled symmetry approach},\ }\href@noop {} {\bibfield  {journal}
  {\bibinfo  {journal}
  {\href{https://arxiv.org/abs/2202.04635}{arXiv:2202.04635}}\ } (\bibinfo
  {year} {2022})}\BibitemShut {NoStop}%
\bibitem [{\citenamefont {Frewer}\ and\ \citenamefont
  {Khujadze}(2021)}]{Frewer21.4}%
  \BibitemOpen
  \bibfield  {author} {\bibinfo {author} {\bibfnamefont {M.}~\bibnamefont
  {Frewer}}\ and\ \bibinfo {author} {\bibfnamefont {G.}~\bibnamefont
  {Khujadze}},\ }\bibfield  {title} {\bibinfo {title} {A critical examination
  of the conformal invariance in the statistical equations of 2{D} turbulent
  scalar fields},\ }\href@noop {} {\bibfield  {journal} {\bibinfo  {journal}
  {\href{https://arxiv.org/abs/2111.02822}{arXiv:2111.02822}}\ } (\bibinfo
  {year} {2021})}\BibitemShut {NoStop}%
\bibitem [{\citenamefont {Frewer}(2018)}]{Frewer18.2}%
  \BibitemOpen
  \bibfield  {author} {\bibinfo {author} {\bibfnamefont {M.}~\bibnamefont
  {Frewer}},\ }\bibfield  {title} {\bibinfo {title} {On new scalings in a
  temporally evolving turbulent plane jet using a different and physical choice
  of equivalence transformations},\ }\href@noop {} {\bibfield  {journal}
  {\bibinfo  {journal}
  {\href{https://hal.archives-ouvertes.fr/hal-01888353v2}{hal:01888353}}\ }
  (\bibinfo {year} {2018})}\BibitemShut {NoStop}%
\bibitem [{\citenamefont {Frewer}\ and\ \citenamefont
  {Khujadze}(2016)}]{Frewer16.3}%
  \BibitemOpen
  \bibfield  {author} {\bibinfo {author} {\bibfnamefont {M.}~\bibnamefont
  {Frewer}}\ and\ \bibinfo {author} {\bibfnamefont {G.}~\bibnamefont
  {Khujadze}},\ }\bibfield  {title} {\bibinfo {title} {On the use of applying
  {L}ie-group symmetry analysis to turbulent channel flow with streamwise
  rotation},\ }\href@noop {} {\bibfield  {journal} {\bibinfo  {journal}
  {\href{https://arxiv.org/abs/1609.08155}{arXiv:1609.08155}}\ } (\bibinfo
  {year} {2016})}\BibitemShut {NoStop}%
\bibitem [{\citenamefont {Frewer}\ \emph
  {et~al.}(2014{\natexlab{a}})\citenamefont {Frewer}, \citenamefont
  {Khujadze},\ and\ \citenamefont {Foysi}}]{Frewer14.2}%
  \BibitemOpen
  \bibfield  {author} {\bibinfo {author} {\bibfnamefont {M.}~\bibnamefont
  {Frewer}}, \bibinfo {author} {\bibfnamefont {G.}~\bibnamefont {Khujadze}},\
  and\ \bibinfo {author} {\bibfnamefont {H.}~\bibnamefont {Foysi}},\ }\bibfield
   {title} {\bibinfo {title} {Is the log-law a first principle result from
  {L}ie-group invariance analysis?},\ }\href@noop {} {\bibfield  {journal}
  {\bibinfo  {journal}
  {\href{https://arxiv.org/abs/1412.3069}{arXiv:1412.3069}}\ } (\bibinfo {year}
  {2014}{\natexlab{a}})}\BibitemShut {NoStop}%
\bibitem [{\citenamefont {Frewer}\ \emph {et~al.}(2017)\citenamefont {Frewer},
  \citenamefont {Khujadze},\ and\ \citenamefont {Foysi}}]{Frewer17}%
  \BibitemOpen
  \bibfield  {author} {\bibinfo {author} {\bibfnamefont {M.}~\bibnamefont
  {Frewer}}, \bibinfo {author} {\bibfnamefont {G.}~\bibnamefont {Khujadze}},\
  and\ \bibinfo {author} {\bibfnamefont {H.}~\bibnamefont {Foysi}},\ }\bibfield
   {title} {\bibinfo {title} {Comment on `{L}ie symmetry analysis of the
  {L}undgren-{M}onin-{N}ovikov equations for multi-point probability density
  functions of turbulent flow'},\ }\href@noop {} {\bibfield  {journal}
  {\bibinfo  {journal}
  {\href{https://arxiv.org/abs/1710.00669}{arXiv:1710.00669}}\ } (\bibinfo
  {year} {2017})}\BibitemShut {NoStop}%
\bibitem [{\citenamefont {Frewer}(2016)}]{Frewer16.4}%
  \BibitemOpen
  \bibfield  {author} {\bibinfo {author} {\bibfnamefont {M.}~\bibnamefont
  {Frewer}},\ }\bibfield  {title} {\bibinfo {title} {Comment on ``{S}ymmetry
  analysis and invariant solutions of the multipoint infinite systems
  describing turbulence"},\ }\href@noop {} {\bibfield  {journal} {\bibinfo
  {journal} {\href{https://doi.org/10.13140/rg.2.2.35698.76480}{ResearchGate}}\
  } (\bibinfo {year} {2016})}\BibitemShut {NoStop}%
\bibitem [{\citenamefont {Khujadze}\ and\ \citenamefont
  {Frewer}(2016)}]{Frewer16.2}%
  \BibitemOpen
  \bibfield  {author} {\bibinfo {author} {\bibfnamefont {G.}~\bibnamefont
  {Khujadze}}\ and\ \bibinfo {author} {\bibfnamefont {M.}~\bibnamefont
  {Frewer}},\ }\bibfield  {title} {\bibinfo {title} {Revisiting the {L}ie-group
  symmetry method for turbulent channel flow with wall transpiration},\
  }\href@noop {} {\bibfield  {journal} {\bibinfo  {journal}
  {\href{https://arxiv.org/abs/1606.08396}{arXiv:1606.08396}}\ } (\bibinfo
  {year} {2016})}\BibitemShut {NoStop}%
\bibitem [{\citenamefont {Frewer}\ \emph {et~al.}(2016)\citenamefont {Frewer},
  \citenamefont {Khujadze},\ and\ \citenamefont {Foysi}}]{Frewer16.1}%
  \BibitemOpen
  \bibfield  {author} {\bibinfo {author} {\bibfnamefont {M.}~\bibnamefont
  {Frewer}}, \bibinfo {author} {\bibfnamefont {G.}~\bibnamefont {Khujadze}},\
  and\ \bibinfo {author} {\bibfnamefont {H.}~\bibnamefont {Foysi}},\ }\bibfield
   {title} {\bibinfo {title} {A note on the notion ``statistical symmetry"},\
  }\href@noop {} {\bibfield  {journal} {\bibinfo  {journal}
  {\href{https://arxiv.org/abs/1602.08039}{arXiv:1602.08039}}\ } (\bibinfo
  {year} {2016})}\BibitemShut {NoStop}%
\bibitem [{\citenamefont {Frewer}\ \emph
  {et~al.}(2015{\natexlab{a}})\citenamefont {Frewer}, \citenamefont
  {Khujadze},\ and\ \citenamefont {Foysi}}]{Frewer15.2}%
  \BibitemOpen
  \bibfield  {author} {\bibinfo {author} {\bibfnamefont {M.}~\bibnamefont
  {Frewer}}, \bibinfo {author} {\bibfnamefont {G.}~\bibnamefont {Khujadze}},\
  and\ \bibinfo {author} {\bibfnamefont {H.}~\bibnamefont {Foysi}},\ }\bibfield
   {title} {\bibinfo {title} {Objections to a {R}eply of {O}berlack et al.},\
  }\href@noop {} {\bibfield  {journal} {\bibinfo  {journal}
  {\href{https://doi.org/10.13140/rg.2.1.1238.2803}{ResearchGate}}\ } (\bibinfo
  {year} {2015}{\natexlab{a}})}\BibitemShut {NoStop}%
\bibitem [{\citenamefont {Frewer}\ \emph
  {et~al.}(2015{\natexlab{b}})\citenamefont {Frewer}, \citenamefont
  {Khujadze},\ and\ \citenamefont {Foysi}}]{Frewer15.1}%
  \BibitemOpen
  \bibfield  {author} {\bibinfo {author} {\bibfnamefont {M.}~\bibnamefont
  {Frewer}}, \bibinfo {author} {\bibfnamefont {G.}~\bibnamefont {Khujadze}},\
  and\ \bibinfo {author} {\bibfnamefont {H.}~\bibnamefont {Foysi}},\ }\bibfield
   {title} {\bibinfo {title} {Comment on ``{S}tatistical symmetries of the
  {L}undgren-{M}onin-{N}ovikov hierarchy"},\ }\href@noop {} {\bibfield
  {journal} {\bibinfo  {journal}
  {\href{https://doi.org/10.1103/PhysRevE.92.067001}{Phys.~Rev.~E}}\ }\textbf
  {\bibinfo {volume} {92}},\ \bibinfo {pages} {067001} (\bibinfo {year}
  {2015}{\natexlab{b}})}\BibitemShut {NoStop}%
\bibitem [{\citenamefont {Frewer}\ \emph
  {et~al.}(2014{\natexlab{b}})\citenamefont {Frewer}, \citenamefont
  {Khujadze},\ and\ \citenamefont {Foysi}}]{Frewer14.1}%
  \BibitemOpen
  \bibfield  {author} {\bibinfo {author} {\bibfnamefont {M.}~\bibnamefont
  {Frewer}}, \bibinfo {author} {\bibfnamefont {G.}~\bibnamefont {Khujadze}},\
  and\ \bibinfo {author} {\bibfnamefont {H.}~\bibnamefont {Foysi}},\ }\bibfield
   {title} {\bibinfo {title} {On the physical inconsistency of a new
  statistical scaling symmetry in incompressible {N}avier-{S}tokes
  turbulence},\ }\href@noop {} {\bibfield  {journal} {\bibinfo  {journal}
  {\href{https://arxiv.org/abs/1412.3061}{arXiv:1412.3061}}\ } (\bibinfo {year}
  {2014}{\natexlab{b}})}\BibitemShut {NoStop}%
\bibitem [{\citenamefont {Sadeghi}\ \emph {et~al.}(2021)\citenamefont
  {Sadeghi}, \citenamefont {Oberlack},\ and\ \citenamefont
  {Gauding}}]{Sadeghi21}%
  \BibitemOpen
  \bibfield  {author} {\bibinfo {author} {\bibfnamefont {H.}~\bibnamefont
  {Sadeghi}}, \bibinfo {author} {\bibfnamefont {M.}~\bibnamefont {Oberlack}},\
  and\ \bibinfo {author} {\bibfnamefont {M.}~\bibnamefont {Gauding}},\
  }\bibfield  {title} {\bibinfo {title} {New symmetry-induced scaling laws of
  passive scalar transport in turbulent plane jets},\ }\href@noop {} {\bibfield
   {journal} {\bibinfo  {journal}
  {\href{https://doi.org/10.1017/jfm.2021.376}{J.~Fluid.~Mech.}}\ }\textbf
  {\bibinfo {volume} {919}},\ \bibinfo {pages} {A5} (\bibinfo {year}
  {2021})}\BibitemShut {NoStop}%
\bibitem [{\citenamefont {Wac{\l}awczyk}\ and\ \citenamefont
  {Oberlack}(2016)}]{Oberlack16}%
  \BibitemOpen
  \bibfield  {author} {\bibinfo {author} {\bibfnamefont {M.}~\bibnamefont
  {Wac{\l}awczyk}}\ and\ \bibinfo {author} {\bibfnamefont {M.}~\bibnamefont
  {Oberlack}},\ }\bibfield  {title} {\bibinfo {title} {Symmetry analysis and
  invariant solutions of the multipoint infinite systems describing
  turbulence},\ }\href@noop {} {\bibfield  {journal} {\bibinfo  {journal}
  {\href{https://doi.org/10.1088/1742-6596/760/1/012038}{J.~Phys.~Conf.~Ser.}}\
  }\textbf {\bibinfo {volume} {760}},\ \bibinfo {pages} {012038} (\bibinfo
  {year} {2016})}\BibitemShut {NoStop}%
\bibitem [{\citenamefont {Oberlack}\ \emph {et~al.}(2022)\citenamefont
  {Oberlack}, \citenamefont {Hoyas}, \citenamefont {Kraheberger}, \citenamefont
  {Alc\'{a}ntara-\'{A}vila},\ and\ \citenamefont {Laux}}]{Oberlack22}%
  \BibitemOpen
  \bibfield  {author} {\bibinfo {author} {\bibfnamefont {M.}~\bibnamefont
  {Oberlack}}, \bibinfo {author} {\bibfnamefont {S.}~\bibnamefont {Hoyas}},
  \bibinfo {author} {\bibfnamefont {S.~V.}\ \bibnamefont {Kraheberger}},
  \bibinfo {author} {\bibfnamefont {F.}~\bibnamefont
  {Alc\'{a}ntara-\'{A}vila}},\ and\ \bibinfo {author} {\bibfnamefont
  {J.}~\bibnamefont {Laux}},\ }\bibfield  {title} {\bibinfo {title} {Turbulence
  statistics of arbitrary moments of wall-bounded shear flows: {A} symmetry
  approach},\ }\href@noop {} {\bibfield  {journal} {\bibinfo  {journal}
  {\href{https://doi.org/10.1103/PhysRevLett.128.024502}{Phys.~Rev.~Lett.}}\
  }\textbf {\bibinfo {volume} {128}},\ \bibinfo {pages} {024502} (\bibinfo
  {year} {2022})}\BibitemShut {NoStop}%
\bibitem [{\citenamefont {Avsarkisov}\ \emph {et~al.}(2014)\citenamefont
  {Avsarkisov}, \citenamefont {Oberlack},\ and\ \citenamefont
  {Hoyas}}]{Avsarkisov14}%
  \BibitemOpen
  \bibfield  {author} {\bibinfo {author} {\bibfnamefont {V.}~\bibnamefont
  {Avsarkisov}}, \bibinfo {author} {\bibfnamefont {M.}~\bibnamefont
  {Oberlack}},\ and\ \bibinfo {author} {\bibfnamefont {S.}~\bibnamefont
  {Hoyas}},\ }\bibfield  {title} {\bibinfo {title} {New scaling laws for
  turbulent {P}oiseuille flow with wall transpiration},\ }\href@noop {}
  {\bibfield  {journal} {\bibinfo  {journal}
  {\href{https://doi.org/10.1017/jfm.2014.98}{J.~Fluid.~Mech.}}\ }\textbf
  {\bibinfo {volume} {746}},\ \bibinfo {pages} {99} (\bibinfo {year}
  {2014})}\BibitemShut {NoStop}%
\bibitem [{\citenamefont {Sadeghi}\ \emph {et~al.}(2018)\citenamefont
  {Sadeghi}, \citenamefont {Oberlack},\ and\ \citenamefont
  {Gauding}}]{Sadeghi18}%
  \BibitemOpen
  \bibfield  {author} {\bibinfo {author} {\bibfnamefont {H.}~\bibnamefont
  {Sadeghi}}, \bibinfo {author} {\bibfnamefont {M.}~\bibnamefont {Oberlack}},\
  and\ \bibinfo {author} {\bibfnamefont {M.}~\bibnamefont {Gauding}},\
  }\bibfield  {title} {\bibinfo {title} {On new scaling laws in a temporally
  evolving turbulent plane jet using {L}ie symmetry analysis and direct
  numerical simulation},\ }\href@noop {} {\bibfield  {journal} {\bibinfo
  {journal} {\href{https://doi.org/10.1017/jfm.2018.625}{J.~Fluid.~Mech.}}\
  }\textbf {\bibinfo {volume} {854}},\ \bibinfo {pages} {233} (\bibinfo {year}
  {2018})}\BibitemShut {NoStop}%
\bibitem [{\citenamefont {Avsarkisov}\ \emph {et~al.}(2021)\citenamefont
  {Avsarkisov}, \citenamefont {Oberlack},\ and\ \citenamefont
  {Hoyas}}]{Avsarkisov21}%
  \BibitemOpen
  \bibfield  {author} {\bibinfo {author} {\bibfnamefont {V.}~\bibnamefont
  {Avsarkisov}}, \bibinfo {author} {\bibfnamefont {M.}~\bibnamefont
  {Oberlack}},\ and\ \bibinfo {author} {\bibfnamefont {S.}~\bibnamefont
  {Hoyas}},\ }\bibfield  {title} {\bibinfo {title} {New scaling laws for
  turbulent {P}oiseuille flow with wall transpiration --
  {C}{O}{R}{R}{I}{G}{E}{N}{D}{U}{M}},\ }\href@noop {} {\bibfield  {journal}
  {\bibinfo  {journal}
  {\href{https://doi.org/10.1017/jfm.2020.1174}{J.~Fluid.~Mech.}}\ }\textbf
  {\bibinfo {volume} {912}},\ \bibinfo {pages} {E2} (\bibinfo {year}
  {2021})}\BibitemShut {NoStop}%
\bibitem [{\citenamefont {Sadeghi}\ \emph {et~al.}(2020)\citenamefont
  {Sadeghi}, \citenamefont {Oberlack},\ and\ \citenamefont
  {Gauding}}]{Sadeghi20.1}%
  \BibitemOpen
  \bibfield  {author} {\bibinfo {author} {\bibfnamefont {H.}~\bibnamefont
  {Sadeghi}}, \bibinfo {author} {\bibfnamefont {M.}~\bibnamefont {Oberlack}},\
  and\ \bibinfo {author} {\bibfnamefont {M.}~\bibnamefont {Gauding}},\
  }\bibfield  {title} {\bibinfo {title} {On new scaling laws in a temporally
  evolving turbulent plane jet using {L}ie symmetry analysis and direct
  numerical simulation -- {C}{O}{R}{R}{I}{G}{E}{N}{D}{U}{M}},\ }\href@noop {}
  {\bibfield  {journal} {\bibinfo  {journal}
  {\href{https://doi.org/10.1017/jfm.2019.985}{J.~Fluid.~Mech.}}\ }\textbf
  {\bibinfo {volume} {885}},\ \bibinfo {pages} {E1} (\bibinfo {year}
  {2020})}\BibitemShut {NoStop}%
\end{thebibliography}%

\end{document}